\documentclass[aps,superscriptaddress,amsmath,amssymb,twocolumn,showpacs,floatfix,english]{revtex4}

\usepackage{url}
\usepackage{bm,bbm}
\usepackage{graphicx}
\usepackage{color}
\usepackage[colorlinks=true, urlcolor=blue, linkcolor=blue, citecolor=blue, pdftex]{hyperref}

\usepackage[english]{babel}

\newcommand{\dagga}{{\phantom{\dagger}}}
\DeclareMathOperator{\Tr}{Tr}

\begin{document}

\title{Competition between spin liquids and valence-bond order in the frustrated spin-$1/2$ Heisenberg
model on the honeycomb lattice}

\author{Francesco Ferrari}
\email[]{frferra@sissa.it}
\affiliation{SISSA-International School for Advanced Studies, Via Bonomea 265, I-34136 Trieste, Italy}

\author{Samuel Bieri}
\affiliation{Institute for Theoretical Physics, ETH Z\"urich, 8099 Z\"urich, Switzerland}

\author{Federico Becca}
\affiliation{Democritos National Simulation Center, Istituto Officina dei Materiali del CNR and
SISSA-International School for Advanced Studies, Via Bonomea 265, I-34136 Trieste, Italy}

\date{\today}

\begin{abstract}
Using variational wave functions and Monte Carlo techniques, we study the antiferromagnetic Heisenberg model with first-neighbor $J_1$ and 
second-neighbor $J_2$ antiferromagnetic couplings on the honeycomb lattice. We perform a systematic comparison of magnetically ordered and 
nonmagnetic states (spin liquids and valence-bond solids) to obtain the ground-state phase diagram. N\'eel order is stabilized for small 
values of the frustrating second-neighbor coupling. Increasing the ratio $J_2/J_1$, we find strong evidence for a continuous transition to 
a nonmagnetic phase at $J_2/J_1 \approx 0.23$. Close to the transition point, the Gutzwiller-projected uniform resonating valence bond state 
gives an excellent approximation to the exact ground-state energy. For $0.23 \lesssim J_2/J_1 \lesssim 0.36$, a gapless $Z_2$ spin liquid with 
Dirac nodes competes with a plaquette valence-bond solid. In contrast, the gapped spin liquid considered in previous works has significantly 
higher variational energy. Although the plaquette valence-bond order is expected to be present as soon as the N\'eel order melts, this ordered 
state becomes clearly favored only for $J_2/J_1 \gtrsim 0.3$. Finally, for $0.36 \lesssim J_2/J_1 \le 0.5$, a valence-bond solid with columnar 
order takes over as the ground state, being also lower in energy than the magnetic state with collinear order. We perform a detailed finite-size 
scaling and standard data collapse analysis, and we discuss the possibility of a deconfined quantum critical point separating the N\'eel 
antiferromagnet from the plaquette valence-bond solid.
\end{abstract}

\pacs{75.10.Jm, 75.10.Kt, 75.40.Mg, 74.40.Kb}

\maketitle

\section{Introduction}\label{sec:intro}

Quantum spin models on two-dimensional frustrated lattices represent important playgrounds where a variety of phases can be attained, emerging 
from zero-point fluctuations. Important examples include gapped and gapless spin liquids or valence-bond states~\cite{Mila2011}. Quantum 
fluctuations are strong when the value of the spin $S$ on each site is small (i.e., for $S=1/2$) and in low spatial dimensionalities (i.e., 
for small coordination number). Furthermore, they are further enhanced in the presence of competing superexchange couplings. In this situation, 
long-range magnetic order can melt even at zero temperature. Then, nonmagnetic ground states can either break some symmetries (e.g., lattice 
translations and/or rotations), leading to a valence-bond solid (VBS), or retain all the symmetries of the Hamiltonian. In the latter case, 
the ground state is known as a quantum spin liquid (or quantum paramagnet). The simplest example in which the combined effect of strong quantum 
fluctuations and spin frustration may give rise to a magnetically disordered ground state is the $J_1\text{-}J_2$ Heisenberg model on the square 
lattice, where both first- and second-neighbor couplings are present. Here, recent numerical calculations predicted a genuine spin-liquid 
behavior for $J_2/J_1 \approx 1/2$. However, it is still unclear whether the spin gap is finite, implying a topological $Z_2$ state, or not, 
thus corresponding to a critical spin liquid~\cite{Jiang2012,Hu2013,Wang2013,Gong2014,Poilblanc2017}. A nonmagnetic phase is expected to appear 
also in the $J_1\text{-}J_2$ model on the triangular lattice, in the vicinity of the classical transition point $J_2/J_1 \approx 1/8$. Also in 
this case, the nature of the ground state is not fully understood, with some calculations supporting gapped excitations (and signatures of 
spontaneously broken lattice point group) and other ones sustaining a gapless spin liquid~\cite{Mishmash2013,Kaneko2014,Zhu2015,Hu2015,Iqbal2016}. 
Finally, a widely studied example in which the ground state does not show magnetic ordering is the Heisenberg model on the kagome lattice. Again, 
the true nature of the ground state is not fully understood as large-scale numerical simulations give conflicting results on the presence of a 
spin gap~\cite{Yan2011,Depenbrock2012,Iqbal2013,He2016,Liao2017}.

All these examples are characterized by an odd number of sites per unit cell and, therefore, according to the Lieb-Schultz-Mattis theorem and its
generalizations~\cite{Lieb1961,Affleck1988,Oshikawa2000,Misguich2002,Hastings2004}, a gapped spectrum implies a degenerate ground state, either
because of some symmetry breaking (leading to a VBS) or due to topological degeneracy (characteristic of $Z_2$ spin liquids). The honeycomb
lattice, with its two sites per unit cell, represents a variation in this respect, and it may therefore show different physical properties
than the previously mentioned cases. The frustrated $J_1\text{-}J_2$ Heisenberg model on this lattice has been investigated by a variety of
analytical and numerical methods, including semiclassical~\cite{Rastelli1979,Fouet2001,Mulder2010}, slave particle~\cite{Wang2010,Lu2011},
and variational approaches~\cite{Clark2011,Mezzacapo2012,Ciolo2014}, coupled-cluster~\cite{Bishop2013} and functional renormalization group
methods~\cite{Reuther2011}, series expansion~\cite{Oitmaa2011}, and exact diagonalization~\cite{Fouet2001,Mosadeq2011,Albuquerque2011}.
Recently, density matrix renormalization group (DMRG) calculations~\cite{Zhu2013,Gong2013} suggested that a plaquette VBS is obtained as soon as
the antiferromagnetic order melts through the frustrating superexchange coupling, i.e., for $J_2 \gtrsim 0.25 J_1$. Furthermore, Ganesh 
{\it et al.}~\cite{Ganesh2013a,Ganesh2013b} claimed the existence of a deconfined quantum critical point, separating the N\'eel from the plaquette 
VBS phase. These DMRG results contradict earlier variational calculations that found an intermediate phase of gapped quantum spin liquid between 
the N\'eel order and the plaquette VBS~\cite{Clark2011}. This spin liquid was identified as the so-called {\it sublattice pairing state} 
(SPS)~\cite{Lu2011,Gong2013,Flint2013}. The SPS was originally motivated by the idea that the half-filled Hubbard model on the honeycomb 
lattice could sustain a gapped spin liquid phase at intermediate values of electron-electron repulsion~\cite{Meng2010}. However, this idea 
eventually turned out to be incorrect~\cite{Sorella2012}.

\begin{figure}
\includegraphics[width=1.0\columnwidth]{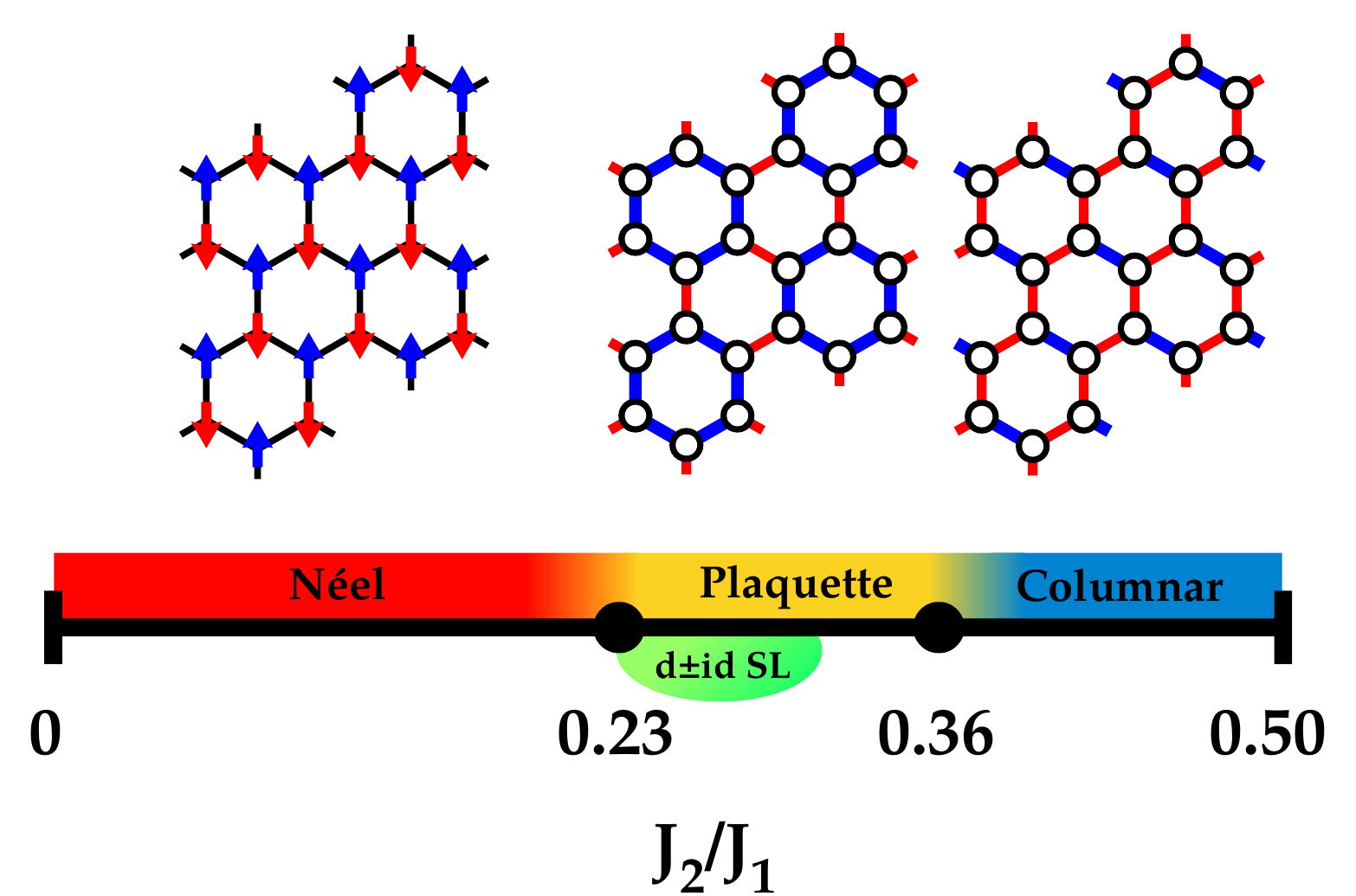}
\caption{\label{fig:phasediagram}
Phase diagram of the spin-$1/2$ $J_1\text{-}J_2$ Heisenberg model on the honeycomb lattice for $0 \le J_2/J_1 \le 0.5$ with schematic illustrations 
of the N\'eel magnetic order, plaquette, and columnar dimer orders. The full dots indicate quantum phase transitions between N\'eel and plaquette 
VBS ($J_2/J_1 \approx 0.23$), and between plaquette and columnar VBS ($J_2/J_1 \approx 0.36$). The region where the $d \pm id$ spin liquid has a 
competitive energy is marked by the green oval.}
\end{figure}

In this paper, we revisit the ground-state phase diagram of the spin-$1/2$ $J_1\text{-}J_2$ Heisenberg model on the honeycomb lattice using 
variational wave functions that can describe both magnetically ordered and disordered phases. As far as the latter are concerned, we perform a 
systematic study of all possible spin liquid {\it Ans\"atze} that have been classified in Ref.~\cite{Lu2011}, including also chiral states.
Moreover, we construct VBS wave functions that are compatible with the previous DMRG simulations (including both plaquette and columnar orders).
Our results show that the N\'eel order melts for $J_2/J_1 \approx 0.23$, in very good agreement with DMRG~\cite{Zhu2013,Gong2013,Ganesh2013b}.
Furthermore, we find that the best spin liquid wave function for $J_2/J_1 \gtrsim 0.23$ is not the gapped SPS as claimed 
earlier~\cite{Clark2011,Gong2013} but instead a symmetric $Z_2$ state with Dirac cones (which is dubbed as $d \pm id$), distinct from all previously 
discussed spin liquid phases. Nonetheless, for $J_2/J_1 \gtrsim 0.3$ we find a substantial energy gain when translation symmetry is broken in the 
variational {\it Ansatz}, suggesting the presence of a plaquette VBS as soon as the N\'eel order melts through spin frustration. Our finite-size 
scaling analysis supports the conclusion of a continuous N\'eel to VBS transition, and may be consistent with the presence of a quantum critical 
point. For even stronger frustration (i.e., $J_2/J_1 \gtrsim 0.36$), a VBS with columnar dimers becomes energetically favored. A sketch of the 
quantum phase diagram is shown in Fig.~\ref{fig:phasediagram}.

The paper is organized as follows: In Sec.~\ref{sec:modelmethods} we give details of the model and the variational wave functions that have been 
employed. In Sec.~\ref{sec:results}, we show the numerical results, and, finally, in Sec.~\ref{sec:conc}, we draw our conclusions.

\section{Model and methods}\label{sec:modelmethods}

\begin{figure}
\includegraphics[width=1.0\columnwidth]{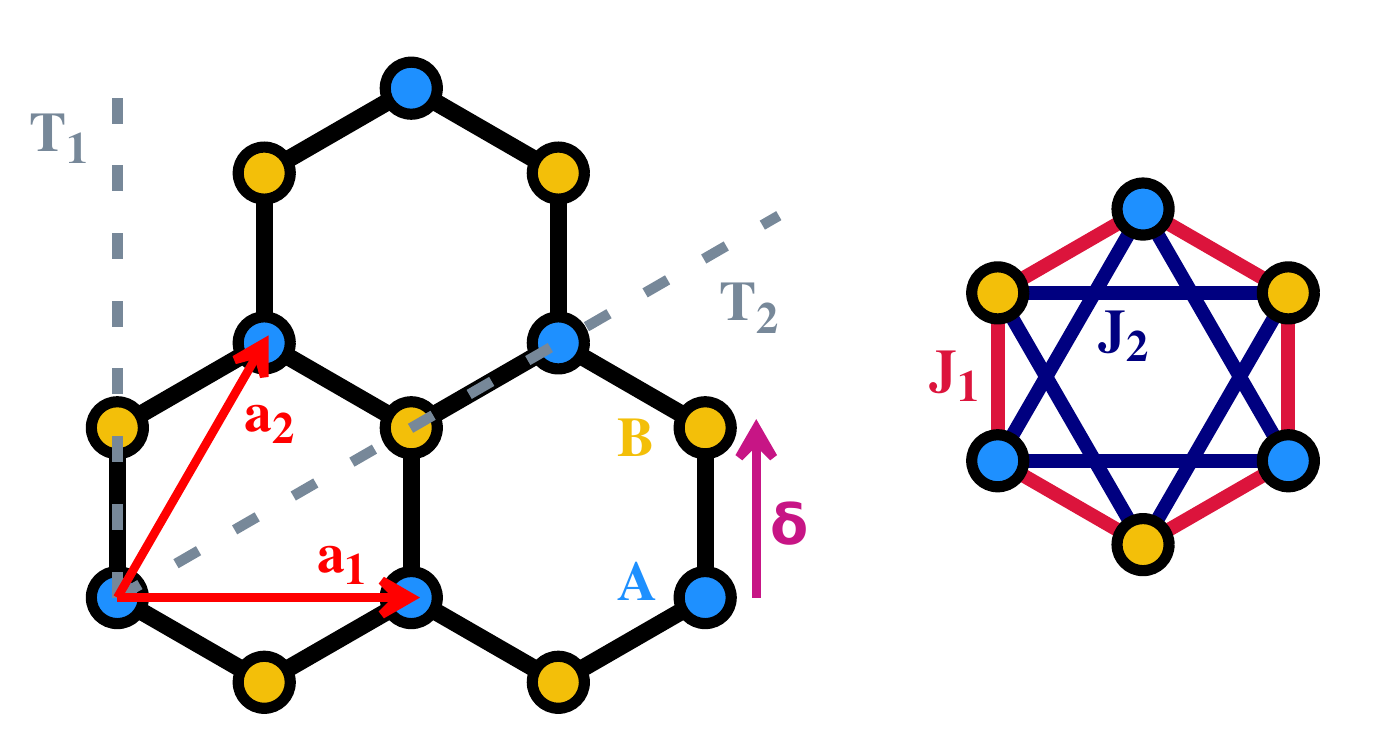}
\caption{\label{fig:honey_j1j2}
The honeycomb lattice is shown on the left: $\mathbf{a}_1$ and $\mathbf{a}_2$ are the primitive vectors of the Bravais lattice. $A$~and $B$ 
denote the two sublattices: $A$-type sites are placed at the origin of the unit cell while $B$-type sites are displaced by $(0,1)$. The dashed 
lines represent the directions of the vectors $\mathbf{T}_1$ and $\mathbf{T}_2$ that define the finite lattice clusters used in the calculations.
A schematic illustration of the interactions in the $J_1\text{-}J_2$ Heisenberg model is shown on the right.}
\end{figure}

The spin-$1/2$ $J_1\text{-}J_2$ Heisenberg model is defined by:
\begin{equation}\label{eq:model}
{\cal H} = J_1 \sum_{\langle i,j \rangle} \mathbf{S}_i \cdot \mathbf{S}_j +
J_2 \sum_{\langle\langle i,j \rangle\rangle} \mathbf{S}_i\cdot \mathbf{S}_j,
\end{equation}
where $\langle i,j \rangle$ and $\langle\langle i,j \rangle\rangle$ denote first- and second-neighbor bonds, respectively (see 
Fig.~\ref{fig:honey_j1j2}). The honeycomb lattice has two sites per unit cell and the underlying Bravais lattice has a triangular structure with 
primitive vectors $\mathbf{a}_1=(\sqrt{3},0)$ and $\mathbf{a}_2=(\sqrt{3}/2,3/2)$. The two sites in the unit cell are labelled by $A$ and $B$: 
the former one is placed in the origin of the cell, while the latter one is displaced by the unit vector $\boldsymbol{\delta} = (0,1)$ (see 
Fig.~\ref{fig:honey_j1j2}). Then, the coordinates of the site $i$ are given by $\mathbf{R}_i = \mathbf{R}_i^0 + \eta_i \boldsymbol{\delta}$, 
where $\mathbf{R}_i^0 = n_i \mathbf{a}_1 + m_i \mathbf{a}_2$, $n_i$ and $m_i$ being integers and the two sites in the unit cell having the same 
$\mathbf{R}_i^0$, and $\eta_i=0$ or $1$. Note that our choice of primitive vectors is such that the first-neighbor distance is equal to $1$. 
For our numerical calculations, we take lattice clusters that are defined by $\mathbf{T}_1 = 2 L \mathbf{a}_2 - L \mathbf{a}_1$ and 
$\mathbf{T}_2=L \mathbf{a}_2 + L \mathbf{a}_1$, thus consisting of $N_s = 6 L^2$ sites (i.e., $3 L^2$ unit cells with two sites each). Periodic 
boundary conditions are imposed on the spin model of Eq.~(\ref{eq:model}).

\begin{figure}
\includegraphics[width=1.0\columnwidth]{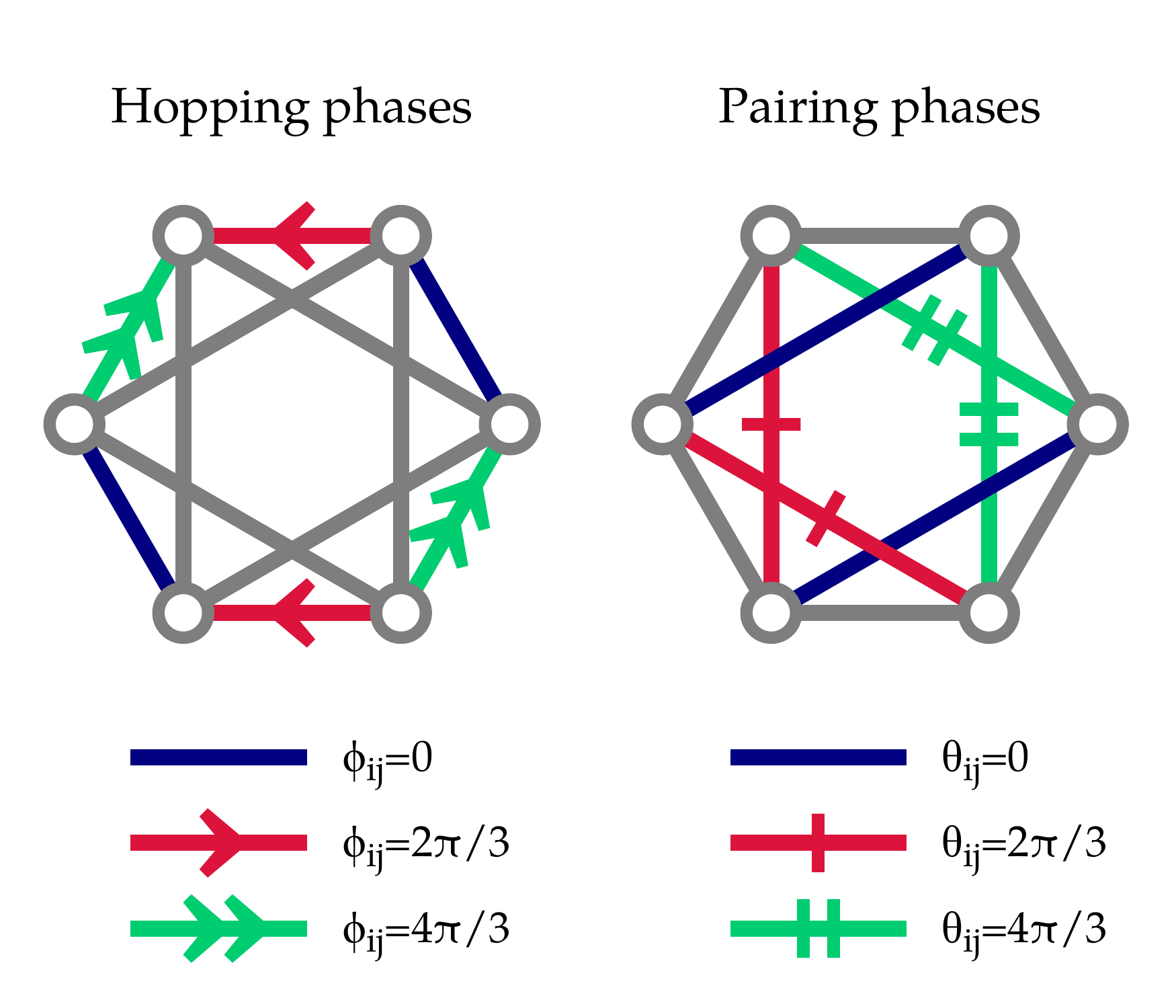}
\caption{\label{fig:ans18}
Schematic illustration of the $d \pm id$ spin liquid state. Here, $\phi_{ij}$ and $\theta_{ij}$ are the complex phases of first-neighbor hopping 
and second-neighbor pairing, respectively. The direction of the arrows ($i\rightarrow j$) indicates the convention of phases for the hopping terms.}
\end{figure}

\begin{figure}
\includegraphics[width=1.0\columnwidth]{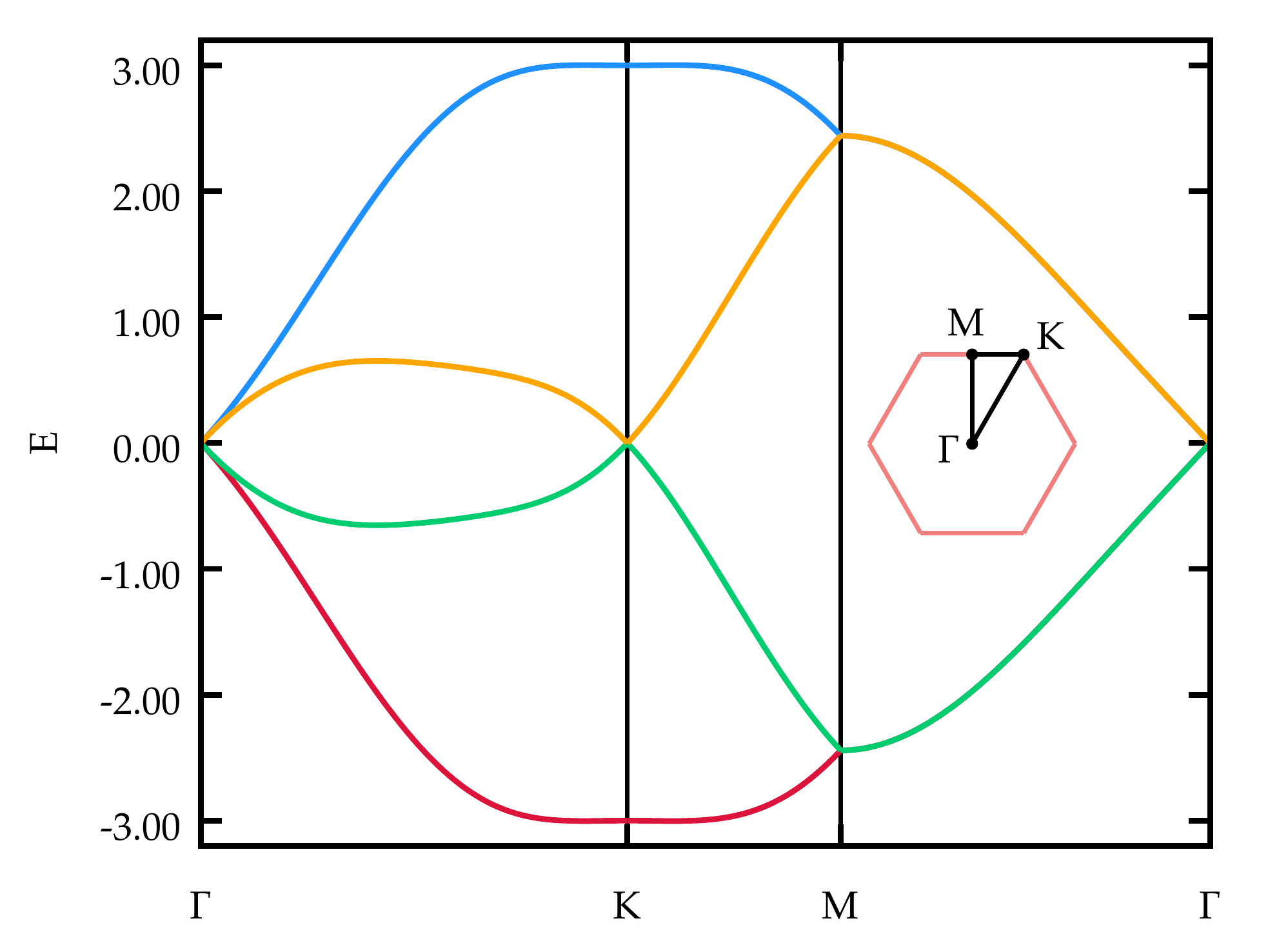}
\caption{\label{fig:ans18_bands}
Mean-field spectrum of the gapless $d \pm id$ pairing state. The energy bands are shown along the path $\Gamma \to K \to M \to \Gamma$ in the 
Brillouin zone (inset). The value of the second-neighbor pairing is $\Delta = 0.35 t$, which is very close to the optimal value obtained for 
$J_2/J_1=0.35$. The bands show Dirac cones located at $\Gamma$ and $K$ points. Note that the two bands are twice degenerate along $M \to \Gamma$.}
\end{figure}

Our results are obtained using variational wave functions constructed from so-called Gutzwiller-projected fermionic states defined as
\begin{equation}\label{eq:wf}
|\Psi\rangle = \mathcal{P}_{S_z^{\rm tot}} \mathcal{J}_z \mathcal{P}_G |\Phi_0\rangle.
\end{equation}
Here, $|\Phi_0\rangle$ is the ground state of suitable quadratic Hamiltonians for auxiliary spinful fermions 
$\{c_{i,\uparrow}^\dagga, c_{i,\downarrow}^\dagga\}$ described below. $\mathcal{P}_G=\prod_i(n_{i,\uparrow}-n_{i,\downarrow})^2$ is the Gutzwiller 
projector that enforces exactly one fermion per site ($n_{i,\sigma} = c_{i,\sigma}^\dagger c_{i,\sigma}^\dagga$), which is needed in order to obtain 
a faithful wave function for the Heisenberg model. $\mathcal{P}_{S_z^{\rm tot}}$ is the projector on the subspace in which the $z$-component of 
the total spin is zero. Finally, $\mathcal{J}_z$ is the spin-spin Jastrow factor:
\begin{equation}\label{eq:jastrow}
\mathcal{J}_z = \exp \left ( -\frac{1}{2} \sum_{i,j} v_{ij} S^z_i S^z_j \right ),
\end{equation}
where the pseudopotential $v_{ij}$ depends on the distance $|\mathbf{R}_i-\mathbf{R}_j|$ (for a translationally invariant system).

Let us now describe in detail the form of the quadratic Hamiltonians that are used to define $|\Phi_0\rangle$.
We will mainly consider two options: one for magnetically ordered, the other for nonmagnetic phases. In the first case, we take:
\begin{equation}\label{eq:hmag}
{\cal H}_{\rm mag} = \sum_{i,j,\sigma} t_{ij} c_{i,\sigma}^\dagger c_{j,\sigma}^\dagga + h.c. +
\sum_i h_i \left ( c_{i,\uparrow}^\dagger c_{i,\downarrow}^\dagga + c_{i,\downarrow}^\dagger c_{i,\uparrow}^\dagga \right ),
\end{equation}
where $t_{ij}$ denotes the hopping amplitude and $h_i$ a (fictitious) magnetic field along the $x$ direction, which is taken to have a
periodic pattern:
\begin{equation}\label{eq:magnetic}
h_j = h \exp \left [ i(\mathbf{Q} \cdot \mathbf{R}_j^0 + \phi_j) \right],
\end{equation}
where $\mathbf{Q}$ is the wave vector that fixes the periodicity, and $\phi_j$ is a sublattice-dependent phase shift. In this work, we consider 
the antiferromagnetic N\'eel phase with $\mathbf{Q}=(0,0)$ and $\phi_j=\eta_j \pi$ (i.e., $\phi_j=0$ for $j \in A$ and $\phi_j = \pi$ for $j \in B$),
and a collinear phase with $\mathbf{Q}=(0, 2\pi/3)$ and $\phi_i=0$. Within this kind of magnetically ordered states, it is very important to take 
into account the spin-spin Jastrow factor in order to introduce transverse spin fluctuations (i.e., spin waves)~\cite{Manousakis1991}. We mention 
in passing that the case $t_{ij}=0$ reduces to a ``bosonic'' (pure Jastrow) state, which has been used by Di~Ciolo {\it et al.} for this 
model~\cite{Ciolo2014}. Interestingly, we find that a nonzero uniform first-neighbor hopping {\it does} provide an energy gain with respect to 
this ``bosonic'' case. This situation is similar to the triangular-lattice antiferromagnet, where a hopping term with Dirac spectrum was also found 
to result in a substantial energy gain~\cite{Iqbal2016}.

In contrast, nonmagnetic phases, such as spin liquids and VBS, can be described by taking:
\begin{eqnarray}\label{eq:hSL}
{\cal H}_{\rm sl} &=& \sum_{i,j,\sigma} t_{ij} c_{i,\sigma}^\dagger c_{j,\sigma}^\dagga +
\sum_{i j} \Delta_{ij} c_{i,\downarrow}^\dagga c_{j,\uparrow}^\dagga + h.c. \nonumber \\
&+& \sum_{i,\sigma} \mu_i c_{i,\sigma}^\dagger c_{i,\sigma}^\dagga + \sum_{i} \zeta_i c_{i,\downarrow}^\dagga c_{i,\uparrow}^\dagga + h.c.,
\end{eqnarray}
where, in addition to the hopping, one introduces singlet pairing terms, $\zeta_i$ and $\Delta_{ij} = \Delta_{ji}$, as well as a chemical potential 
$\mu_i$. Within this framework, a classification of distinct spin-liquid phases can be obtained through the so-called projective symmetry group 
(PSG) analysis~\cite{Wen2002,Bieri2016}. From a variational perspective, the PSG provides a recipe for constructing symmetric spin-liquid wave 
functions through specific {\it Ans\"atze} for the Hamiltonian~(\ref{eq:hSL}). The simplest {\it Ansatz} is given by a first neighbor hopping 
($t_{ij}=t$) and no pairing terms ($\Delta_{ij} = \zeta_i = 0$). This is the {\it uniform resonating valence bond} (uRVB) state, which is a $U(1)$ 
state with Dirac cones at the corners of the hexagonal Brillouin zone. By performing a PSG classification, Lu and Ran~\cite{Lu2011} found $24$ 
symmetric $Z_2$ spin liquids that are continuously connected to this uRVB (i.e., that can be obtained from uRVB by adding further hopping and/or 
pairing terms). Among those states, the presence of the gapped SPS was emphasized. The SPS {\it Ansatz} is characterized by a uniform first-neighbor 
hopping $t$ and a complex second-neighbor pairing with opposite phases on $A$-$A$ and $B$-$B$ links, i.e., ${\Delta_{ij}^{AA}=\Delta e^{i\theta}}$ 
and ${\Delta_{ij}^{BB}=\Delta e^{-i\theta}}$. Such a state is always gapped if ${\Delta \neq 0}$ and ${\theta \neq \pi/2}$. In principle, the PSG 
classification also allows an on-site pairing with opposite phases on the two sublattices, i.e., ${\zeta_i^A=\zeta e^{i\phi}}$, 
${\zeta_i^B=\zeta e^{-i\phi}}$. In agreement with previous studies~\cite{Clark2011,Gong2013}, we find that the SPS {\it Ansatz} has a lower 
variational energy than the uRVB state for $J_2/J_1 \gtrsim 0.25$. The actual value of $\theta$ can be set to zero since the variational energy 
does not change appreciably for $\theta \lesssim \pi/4$. However, here we find another {\it gapless} spin liquid (i.e., number~$18$ in Table~I of 
Ref.~\cite{Lu2011}) that has an even lower energy than the SPS wave function and represents the best $Z_2$ state among those classified within the 
fermionic PSG. We adopt a natural gauge in which this spin liquid {\it Ansatz} has first-neighbor hopping $t_{ij} = t e^{i\phi_{ij}}$ and 
second-neighbor pairing $\Delta_{ij} = \Delta e^{i\theta_{ij}}$ with complex phases as given in Fig.~\ref{fig:ans18}, a convention that differs
from the original PSG solution of Ref.~\cite{Lu2011}. Since $\Delta_{ij}$ has a $d_{x^2-y^2}+i d_{xy}$ phase winding on the triangular lattice of 
$A$ sites and $d_{x^2-y^2}-id_{xy}$ on the $B$ sublattice, we call this new state $d \pm id$. For $\Delta=0$, the $d \pm i d$ state reduces to 
uRVB, while for $t=0$, it is two copies of the {\it quadratic band touching} state that has been discussed in Ref.~\cite{Mishmash2013} for the 
triangular lattice. For finite $\Delta$, the fermionic mean-field energy bands show Dirac nodes at the center and at the corners of the Brillouin 
zone (see Fig.~\ref{fig:ans18_bands}). Note that, despite the presence of complex hopping and pairing terms, both the SPS and the $d \pm id$ states 
do not break time-reversal symmetry (or any other lattice symmetry) once the wave function is Gutzwiller-projected to the physical spin Hilbert 
space (see Appendix~\ref{app:PSG} for its projective symmetries). Beyond fully symmetric phases, we also looked for potential chiral spin liquids 
as outlined in Ref.~\cite{Bieri2016}. However, we do not find any indication for such ground states in the present model.

\begin{figure}
\includegraphics[width=1.0\columnwidth]{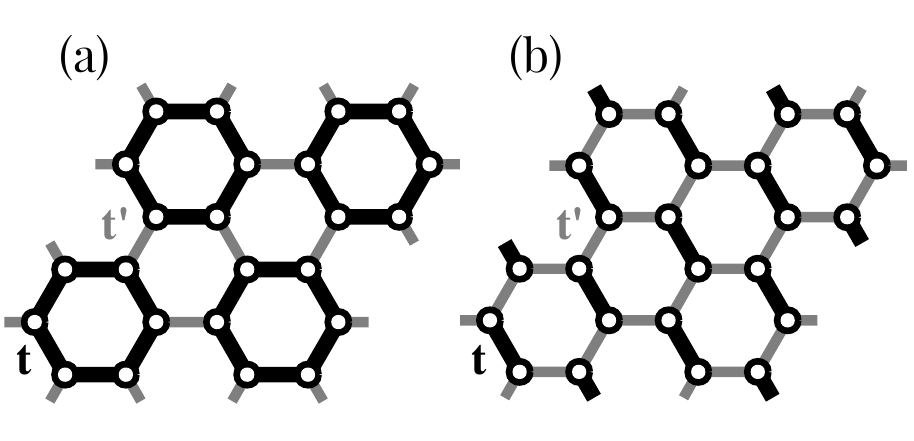}
\caption{\label{fig:plaquette}
Patterns of the first-neighbor hoppings in the quadratic Hamiltonian~(\ref{eq:hSL}) as found in the plaquette VBS (a) and in the columnar VBS (b).}
\end{figure}

\begin{figure}
\includegraphics[width=1.0\columnwidth]{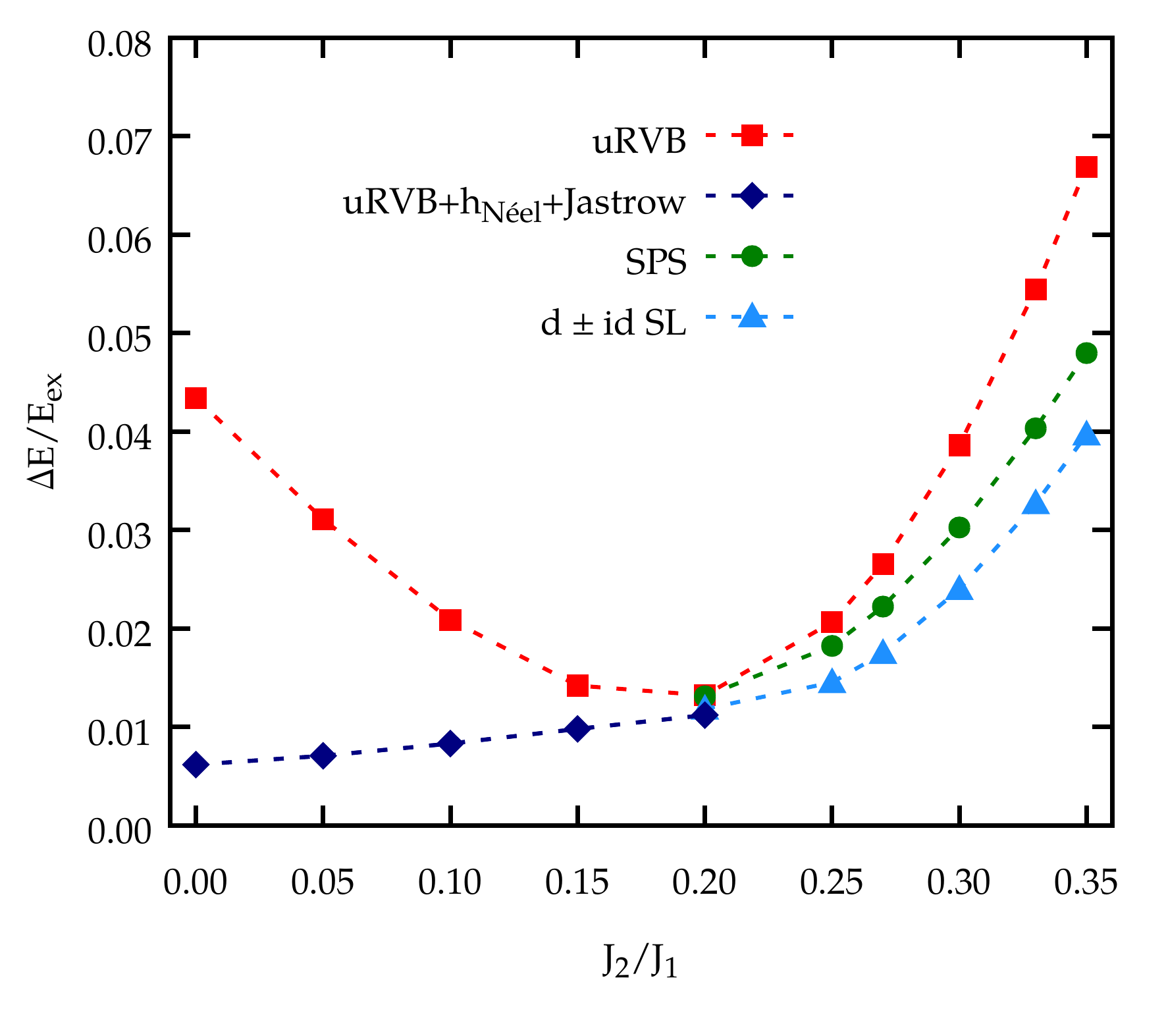}
\caption{\label{fig:accuracy}
Accuracy of the variational energy for different wave functions on the $24$-site cluster. Here, $\Delta E$ is the difference between the
variational ($E_{var}$) and the exact ground-state energy ($E_{ex}$).}
\end{figure}

Using the Hamiltonian of Eq.~(\ref{eq:hSL}), we can also construct wave functions with VBS order. This can be achieved by allowing a translation 
and/or rotation symmetry breaking in the hopping $t_{ij}$ and/or in the pairing $\Delta_{ij}$ parameters. Here, we consider two possibilities 
which are motivated by recent DMRG results~\cite{Zhu2013,Gong2013,Ganesh2013b}. These are obtained by considering two different first-neighbor 
hoppings $t$ and $t^\prime$, forming ``strong'' and ``weak'' plaquettes or columnar dimers, see Fig.~\ref{fig:plaquette}. In both cases, a 
remarkable improvement in variational energy is achieved by adding a (uniform) second-neighbor pairing with $d \pm id$ symmetry, as well as 
including the corresponding complex phases for the dimerized first-neighbor hoppings (Fig.~\ref{fig:ans18}). These are rare examples of clear 
VBS instabilities in frustrated two-dimensional Heisenberg models using Peierls-type mean-field parameters in Gutzwiller-projected wave functions 
(see, e.g., Ref.~\cite{Iqbal2012}).

Finally, we would like to emphasize that, in order to calculate observables (e.g., the variational energy, or any correlation function) in the 
state of Eq.~(\ref{eq:wf}), Monte Carlo sampling is needed, since an analytic treatment is not possible in two spatial dimensions. The optimal 
variational parameters (including the ones defining the quadratic Hamiltonian and the Jastrow pseudo potential), for each value of the ratio 
$J_2/J_1$, can be obtained using the stochastic reconfiguration technique~\cite{Sorella2005}.

\section{Results}\label{sec:results}

In the following, we show the numerical results obtained by the variational approach described in the previous section.

\subsection{Accuracy of the wave functions}\label{sec:acc}

Let us first discuss the accuracy of the optimized variational energy for various states on a small lattice cluster with $24$ sites (i.e., $L=2$) 
for which exact diagonalization is available. In Fig.~\ref{fig:accuracy}, we present the results for the uRVB state (with only first-neighbor
hopping), the N\'eel state (also including the fictitious magnetic field $h_i$ and the spin-spin Jastrow factor), the SPS {\it Ansatz} (with
second-neighbor pairing and $\theta=0$), and the $d \pm id$ state. First of all, starting from the unfrustrated limit with $J_2=0$, the accuracy 
of the uRVB state clearly improves until $J_2/J_1 \approx 0.2$. Then the energy rapidly deteriorates when $J_2/J_1$ is further increased. For 
$J_2/J_1 \lesssim 0.2$, the best variational state is given by including N\'eel order with $\mathbf{Q}=(0,0)$ and $\phi_i=\eta_i \pi$. In this regime, 
the strength of the magnetic field $h$ in Eq.~(\ref{eq:magnetic}) decreases as $J_2/J_1$ increases, and it goes to zero for $J_2/J_1 \gtrsim 0.2$. 
When $h=0$, only a marginal energy gain with respect to the uRVB state is obtained, due to a (small) spin-spin Jastrow factor. For this reason, 
the results for the N\'eel state are not reported for $J_2/J_1 > 0.2$. In contrast, an energy gain is found in this regime by allowing a pairing 
term in Eq.~(\ref{eq:hSL}). Here, both the SPS and the $d \pm id$ {\it Ans\"atze} give a lower variational energy than the simple uRVB. We emphasize 
that the $d \pm id$ wave function represents the best fermionic state among the $24$ $Z_2$ spin liquids listed in Ref.~\cite{Lu2011}. On the small 
cluster considered, there is no significant energy gain by allowing VBS order on top of the $d \pm id$ state for $J_2/J_1 \lesssim 0.35$.

\begin{figure}
\includegraphics[width=1.0\columnwidth]{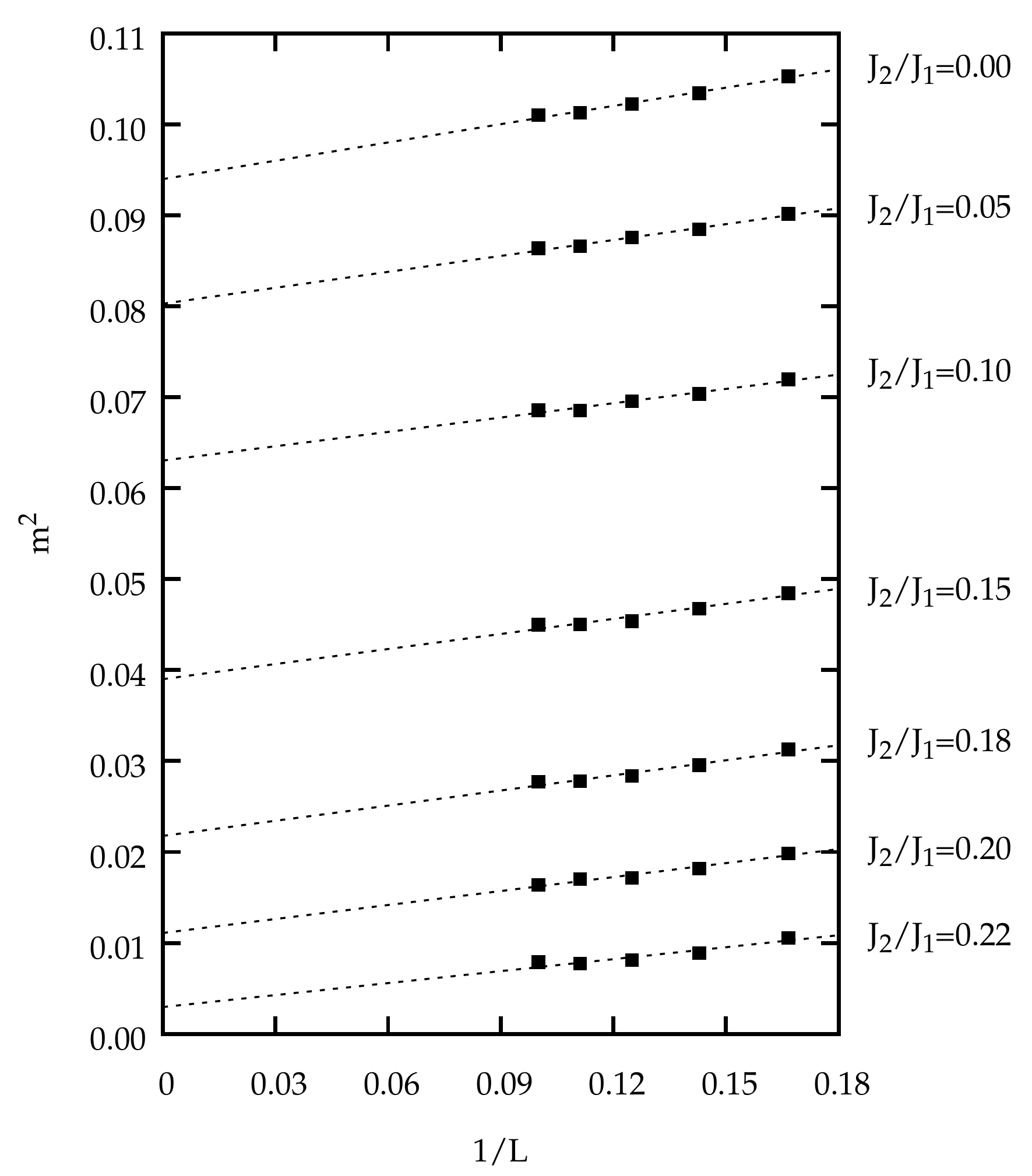}
\caption{\label{fig:magn_scaling}
Finite-size scaling of the squared magnetization $m^2$, Eq.~\eqref{eq:magnetization}, for different values of $J_2/J_1$. For $J_2=0$ the optimal 
value of the fictitious magnetic field is $h/t \approx 0.32$.}
\end{figure}

\begin{figure}
\includegraphics[width=1.0\columnwidth]{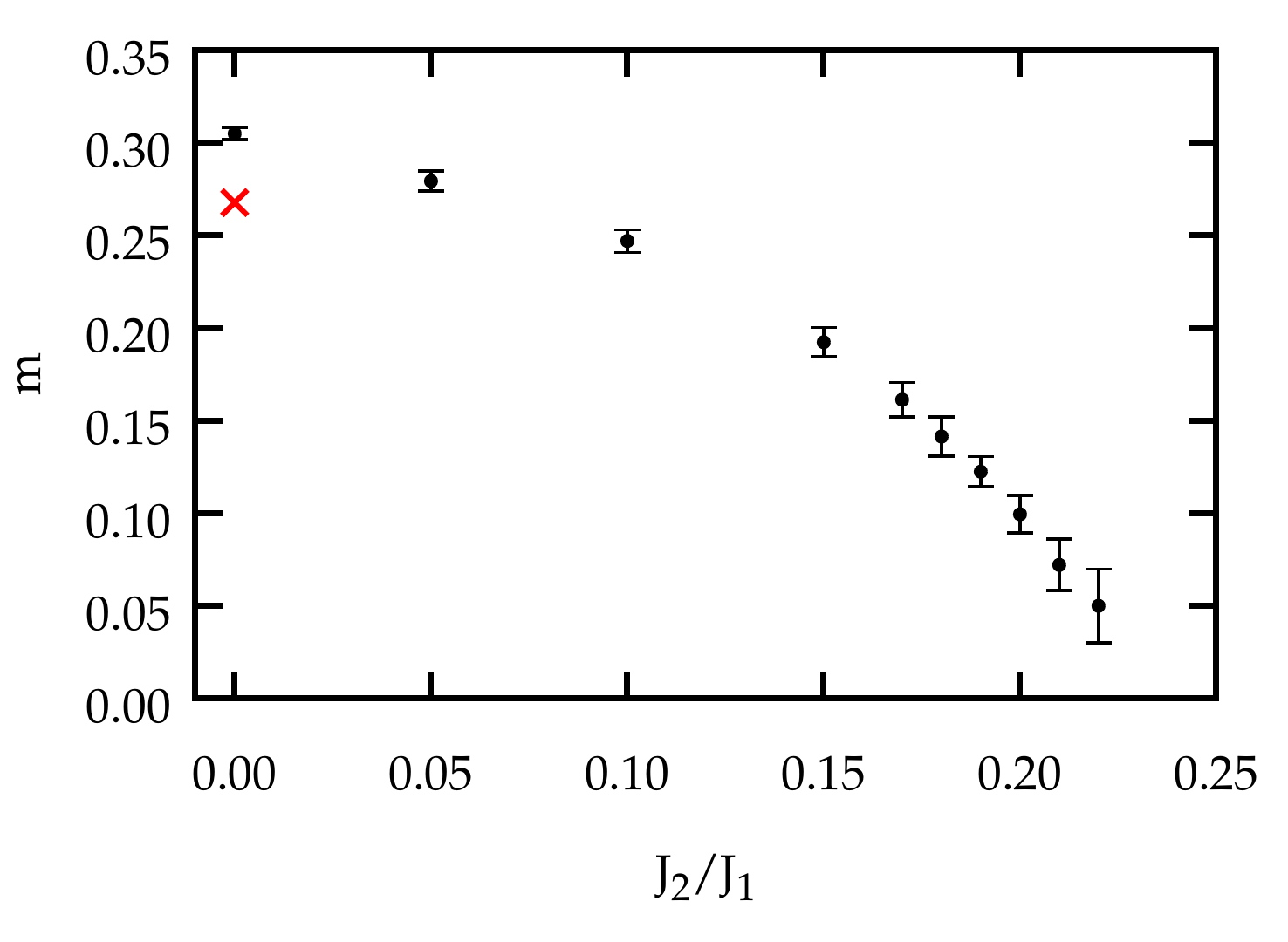}
\caption{\label{fig:magn_vs_J2}
Thermodynamic limit of the magnetization $m$ as a function of $J_2/J_1$. The result from quantum Monte Carlo of Ref.~\cite{Castro2006}
for $J_2=0$ is shown for comparison (red cross). The classical value is $m = 0.5$.
}
\end{figure}

\subsection{The N\'eel phase}

In order to draw the ground-state phase diagram, we focus on the N\'eel phase and perform a finite-size scaling of the magnetization, which is 
obtained from the expectation value of the spin-spin correlation at the maximal distance
\begin{equation}\label{eq:magnetization}
m^2 = \lim_{|i-j| \to \infty} \langle \mathbf{S}_i \cdot \mathbf{S}_j \rangle,
\end{equation}
in the variational state $|\Psi\rangle$. The results for $0 \le J_2/J_1 \le 0.22$ are reported in Fig.~\ref{fig:magn_scaling} for $L$ ranging from 
$6$ to $10$ (i.e., up to $600$ sites). The thermodynamic extrapolation of the magnetization $m$ is shown in Fig.~\ref{fig:magn_vs_J2}. The expected 
$1/L$ corrections are correctly reproduced by the spin-spin Jastrow factor, which is able to introduce the relevant low-energy fluctuations on top 
of the classical order parameter that is generated by the magnetic field $h$ of Eq.~(\ref{eq:magnetic}). The thermodynamic value of the staggered 
magnetization vanishes for $J_2/J_1 \approx 0.23$ (see also the discussion in Sec.~\ref{sec:conc}), in good agreement with previous DMRG 
calculations~\cite{Zhu2013,Gong2013,Ganesh2013b}. We remark that the value $J_2/J_1 \approx 0.23$ is larger than the one obtained in the classical 
limit (i.e., $J_2/J_1 = 1/6$), indicating that quantum fluctuations favor collinear magnetic order over generic coplanar spirals (which represent 
the classical ground state for $J_2/J_1>1/6$). Comparison with exact quantum Monte Carlo calculations, which are only possible in the unfrustrated 
case $J_2=0$~\cite{Castro2006}, further substantiates the accuracy of the N\'eel wave function on large systems, see Fig.~\ref{fig:magn_vs_J2}. 
Even though a direct inspection of our numerical results cannot exclude a first-order transition at $J_2/J_1 \approx 0.23$, a detailed finite-size 
scaling analysis based on data collapse suggests that the transition between the N\'eel and the nonmagnetic phase is continuous (see below).

\subsection{The nonmagnetic phase}

Increasing the ratio $J_2/J_1$, the N\'eel order melts and the natural expectation is that a nonmagnetic phase is stabilized by quantum fluctuations.
Nonetheless, we cannot exclude that magnetic states with {\it incommensurate} spirals are favored instead, as it happens in the classical limit 
for $J_2/J_1>1/6$. In any numerical calculation that considers finite clusters, it is very difficult to assess states with large periodicity or 
with pitch vectors that are not allowed by the finite cluster geometry. Therefore, we will not consider the possibility of incommensurate spiral 
orders here, and we restrict ourselves to states with collinear order, i.e., the one with $\mathbf{Q}=(0,2\pi/3)$ and $\phi_i=0$. This restriction 
is justified by recent variational Monte Carlo results showing that collinear (or short-period spirals) may prevail over generic states with long 
periodicity~\cite{Ciolo2014}. As far as the nonmagnetic states are concerned, we consider the ones that can be constructed with the help of the 
Hamiltonian~(\ref{eq:hSL}). For these cases, we do not include the spin-spin Jastrow factor~(\ref{eq:jastrow}), since this term would break SU(2) 
spin rotation symmetry (in any case, the inclusion of a Jastrow factor only leads to minor energy gains). 

\begin{figure}
\includegraphics[width=1.0\columnwidth]{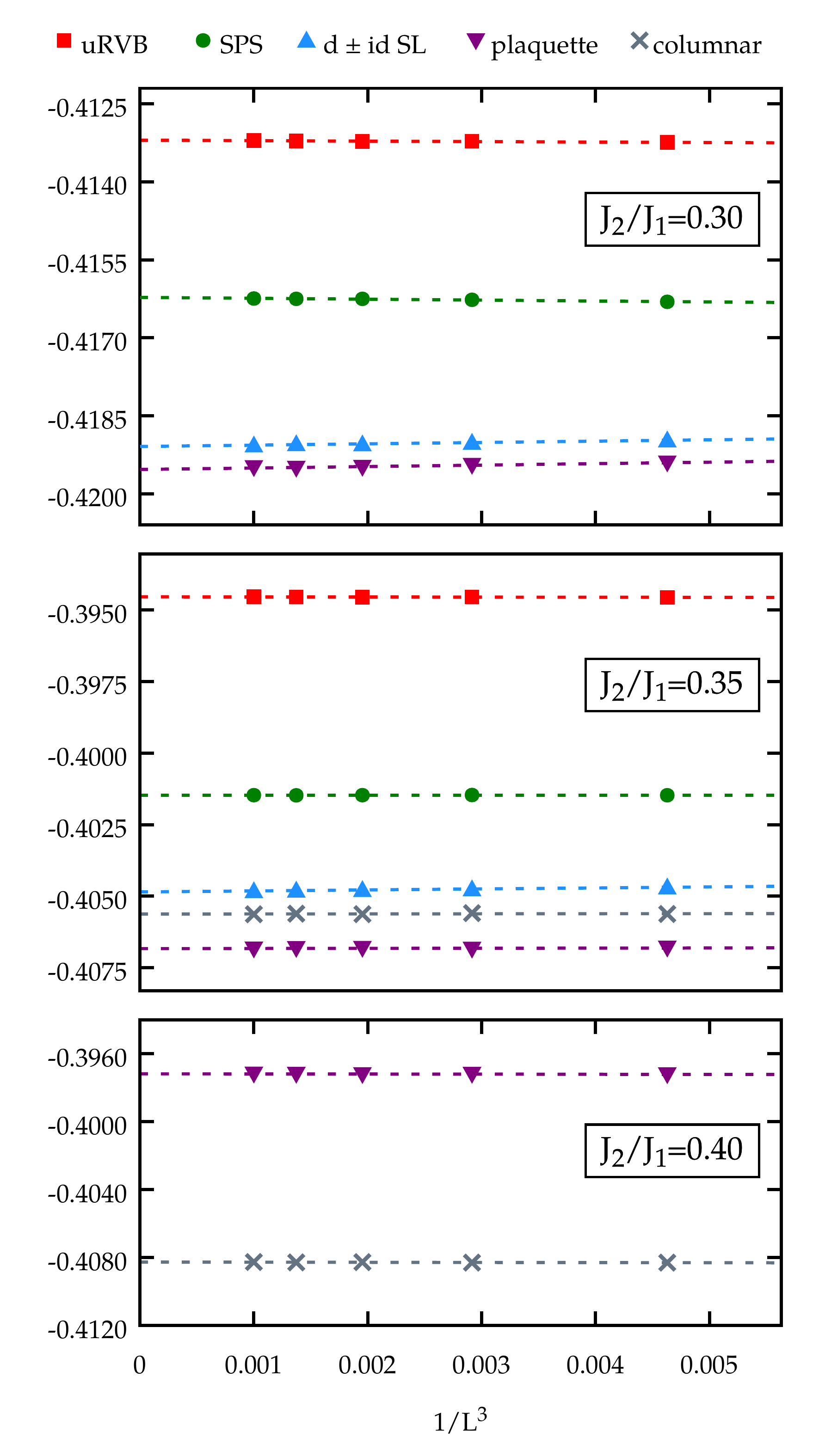}
\caption{\label{fig:SLscaling}
Finite-size scaling analysis of the variational energy of different wave functions for $J_2/J_1=0.3$, $0.35$, and $0.4$. Statistical errors
are smaller than the symbol size. The uRVB, SPS, and $d \pm id$ states are reported only for $J_2/J_1=0.3$ and $0.35$; for $J_2/J_1=0.4$,
their energies are much higher than the ones of plaquette and columnar dimer VBS states. The dashed lines are the fitting functions used for the
extrapolation to the thermodynamic limit. For the $d \pm id$ state, the optimal values of the parameters are $\Delta/t \approx 0.31$ and 
$\Delta/t \approx 0.36$ for $J_2/J_1=0.3$ and $0.35$, respectively. For the plaquette state the parameters range from $\Delta/t \approx 0.31$ 
and $t^\prime/t \approx 0.90$ for $J_2/J_1=0.3$ to $\Delta/t \approx 0.38$ and $t^\prime/t \approx 0.62$ for $J_2/J_1=0.4$. Finally, for the 
columnar state we get $\Delta/t \approx 0.34$ ($\Delta/t \approx 0.37$) and $t^\prime/t \approx 0.45$ ($t^\prime/t \approx 0.20$) for 
$J_2/J_1=0.35$ ($J_2/J_1=0.4$).}
\end{figure}

\begin{figure}
\includegraphics[width=1.0\columnwidth]{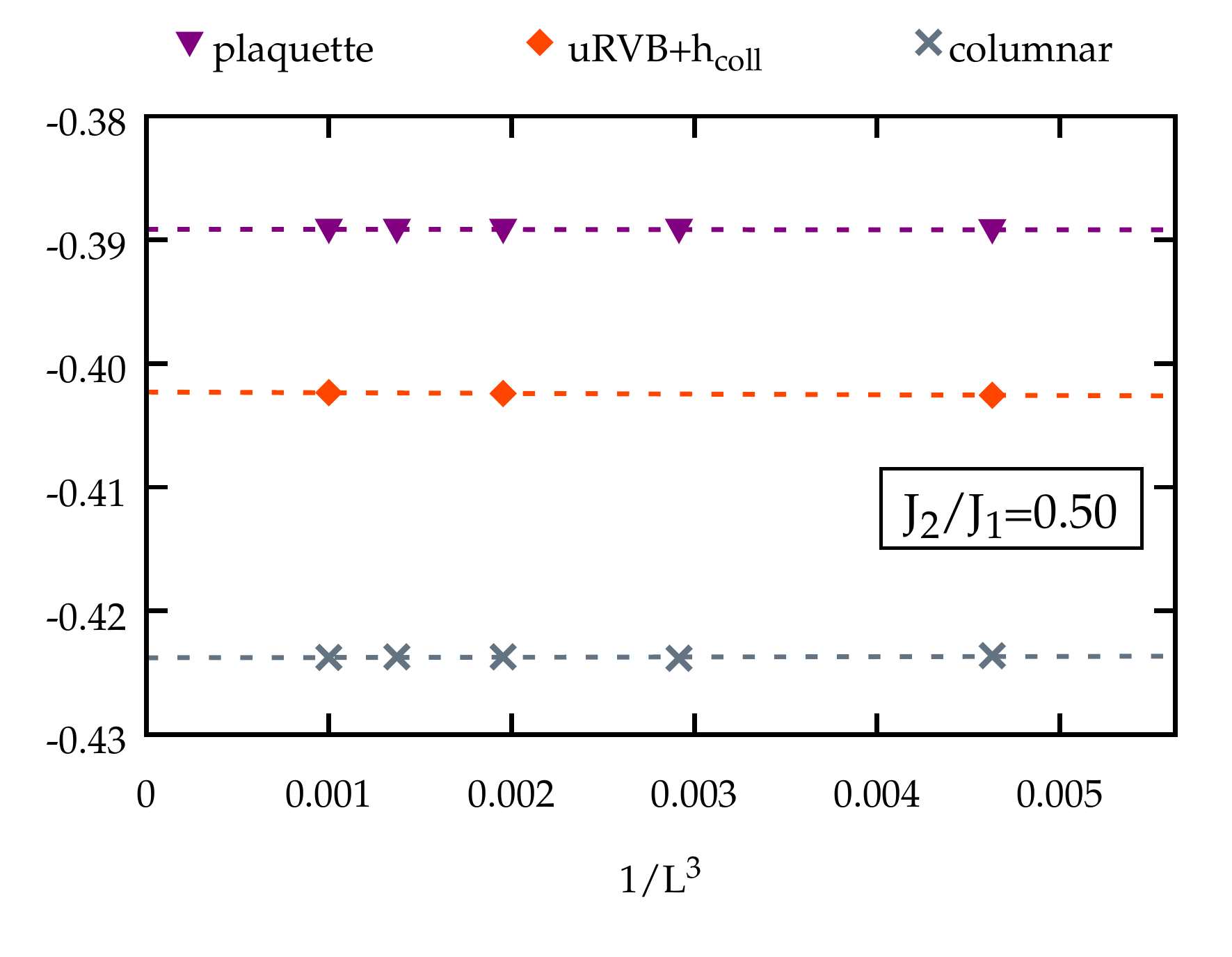}
\caption{\label{fig:collplaq}
Finite-size scaling analysis of the variational energy of VBS and collinear magnetic states at $J_2/J_1=0.5$. The optimal values for the 
plaquette state are $\Delta/t \approx 0.40$ and $t^\prime/t\approx0.20$. For the columnar state $t^\prime/t$ is essentially zero, while 
$\Delta/t \approx 0.37$. The fictitious magnetic field of the collinear state has optimal value $h/t \approx 0.98$.}
\end{figure}

In Fig.~\ref{fig:SLscaling}, we report the finite-size scaling of the energies for $J_2/J_1=0.3$, $0.35$, and $0.4$. Various variational wave functions 
are reported, since the uRVB is unstable when adding pairing terms or allowing a translation symmetry breaking in the quadratic Hamiltonian.
First of all, the SPS {\it Ansatz} gives a size-consistent improvement with respect to the uRVB state in both cases. Our calculations are shown 
for $\theta=0$. In addition to the second-neighbor pairing, the symmetry-allowed nonzero on-site pairing leads to a {\it gapless} mean-field 
spectrum, spoiling the gapped nature of the SPS {\it Ansatz}. However, this variational freedom does not give an appreciable energy gain for the 
values of $J_2/J_1$ considered here. The best spin-liquid wave function, among the $24$ possibilities listed in Ref.~\cite{Lu2011}, is the 
$d \pm id$ state discussed in Sec.~\ref{sec:modelmethods}. But most strikingly, the lowest-energy state in this regime has plaquette VBS order, 
where the first-neighbor hoppings exhibit the pattern shown in Fig.~\ref{fig:plaquette}(a). Here, the presence of a second-neighbor pairing with 
$d \pm id$ symmetry gives a significant improvement in the variational energy, but the stabilization of a plaquette state is already observed 
using first-neighbor hopping only. Its energy gain with respect to the uniform $d \pm id$ {\it Ansatz} clearly increases with increasing $J_2/J_1$, 
being approximately $5 \times 10^{-4}J_1$ for $J_2/J_1=0.3$ and $2 \times 10^{-3}J_1$ for $J_2/J_1=0.35$ (see Fig.~\ref{fig:SLscaling}).

Further increasing $J_2/J_1$, a different VBS with columnar order wins over the plaquette VBS, see Fig.~\ref{fig:SLscaling}. The corresponding 
pattern of first-neighbor hoppings is shown in Fig.~\ref{fig:plaquette}(b). Again, a second-neighbor $d \pm id$ pairing gives a substantial energy 
gain, allowing us to obtain a stable optimization of the columnar order. The fact that both columnar and plaquette states can be stabilized, even 
when their respective energy is higher than the one of the competitor, strongly suggests that the transition between these two VBS phases is first 
order. Based on the calculation of variational energies on relatively large clusters, our estimation of the transition point is $J_2/J_1 \approx 0.36$
(in remarkably good agreement with DMRG~\cite{Zhu2013,Ganesh2013b}).

Finally, we briefly discuss the possible emergence of magnetic order close to $J_2/J_1=0.5$. Unfortunately, the pitch vector of the relevant
magnetic state that is found at the classical and semiclassical levels varies continuously with $J_2/J_1$~\cite{Fouet2001,Mulder2010}. This fact 
makes it impossible to determine the best spiral state on finite clusters. However, for $J_2/J_1=0.5$, the classical state that is selected by 
quantum fluctuations is relatively simple, having collinear order. More specifically, it has spins that are antiferromagnetically aligned on two 
out of the three first-neighbor directions, and ferromagnetically aligned on the third direction. There are three inequivalent possibilities for 
this ordering (corresponding to the choice of the ferromagnetic bond) and, therefore, this state breaks rotation symmetry (similar to the 
$J_1\text{-}J_2$ model on the square lattice for $J_2/J_1>0.5$~\cite{Chandra1990}). In the following, we compare the VBS and the collinear magnetic 
state for $J_2/J_1 = 0.45$ and $0.5$. We take the best VBS {\it Ansatz}, which is given by the columnar state (including the $d \pm id$ pairing), 
and a magnetically ordered wave function, which is constructed using Eq.~(\ref{eq:hmag}) with $\mathbf{Q}=(0,2\pi/3)$ and $\phi_i=0$. The results 
of the finite-size scaling of the energies are shown in Fig.~\ref{fig:collplaq} for $J_2/J_1=0.5$ (similar results are obtained for $J_2/J_1=0.45$).
In this regime, the VBS {\it Ansatz} overcomes the collinear state with a remarkable energy gain. Therefore, we can safely affirm that, for 
$0.36 \lesssim J_2/J_1 \le 0.5$, the best variational wave function exhibits VBS order. These results are in agreement with previous 
studies~\cite{Clark2011,Ganesh2013b}, which detected signatures of rotation-symmetry breaking, and suggested the existence of a dimerized phase for 
large values of $J_2/J_1$.

\subsection{N\'eel to VBS transition: finite-size scaling analysis}

\begin{figure}
\includegraphics[width=1.0\columnwidth]{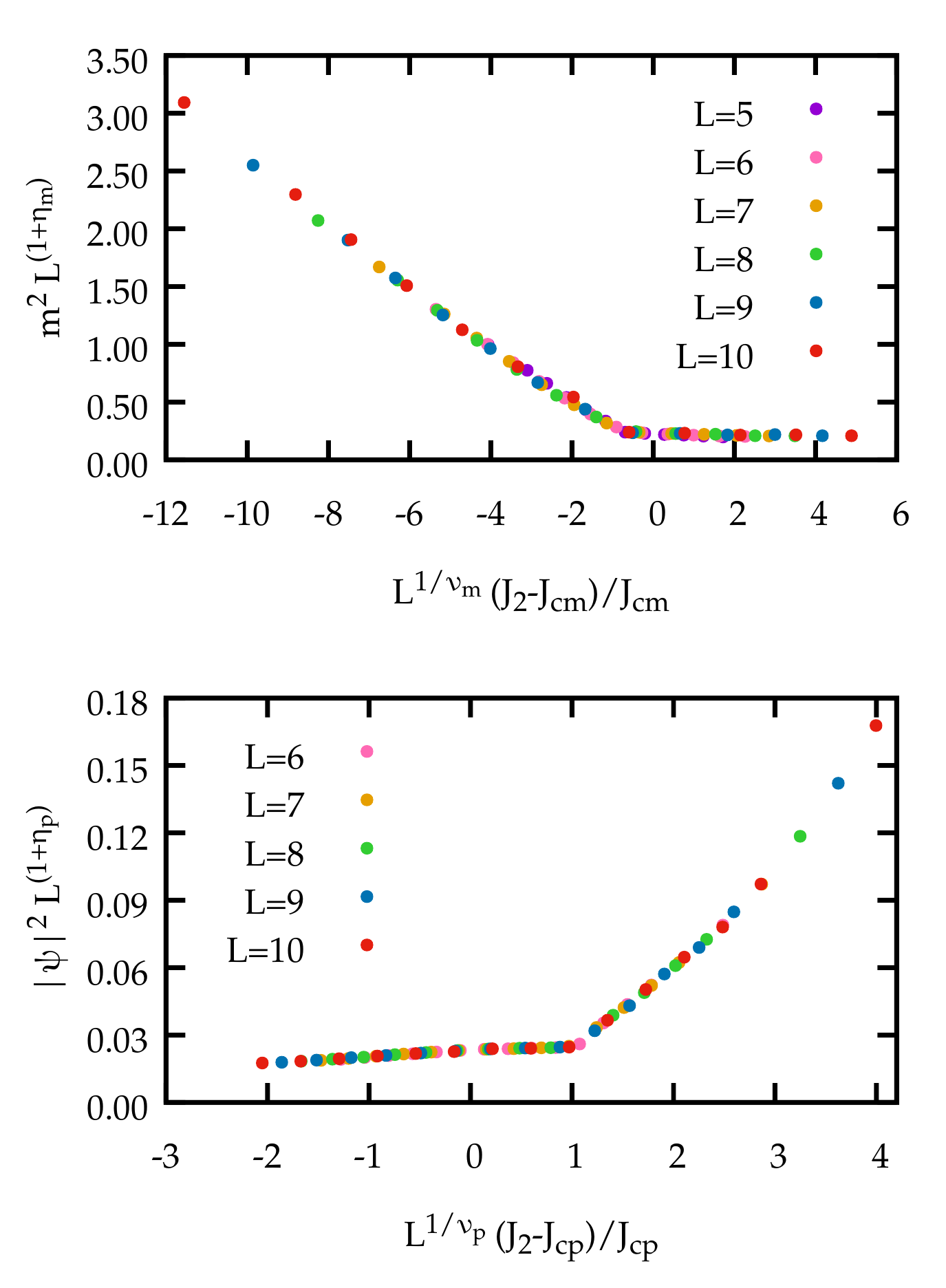}
\caption{\label{fig:FSS}
Finite-size scaling collapse of the data of the antiferromagnetic (above) and plaquette (below) order parameters.}
\end{figure}

In this last section, we briefly discuss the possibility for the N\'eel to VBS transition to be an example of the so-called {\it deconfined
quantum criticality}~\cite{Senthil2004a,Senthil2004b} as suggested by Ganesh {\it et al.}~\cite{Ganesh2013a,Ganesh2013b}. We compute both the 
magnetization [see Eq.~(\ref{eq:magnetization})] and the VBS order parameter:
\begin{equation}
\psi = \frac{1}{N} \sum_{i \in A} \langle D_i \rangle e^{-i\frac{2\pi}{3}(n_i-m_i)},
\end{equation}
where
\begin{equation}
D_i = S^z_i S^z_{i+x} + S^z_i S^z_{i+y} e^{i\frac{2\pi}{3}} + S^z_i S^z_{i+z} e^{-i\frac{2\pi}{3}}.
\end{equation}
Here, site $i$ has coordinates $\mathbf{R}_i=n_i \mathbf{a}_1 + m_i \mathbf{a}_2$ (belonging to sublattice $A$), while sites $i+x$, $i+y$, and $i+z$ 
have coordinates $\mathbf{R}_i-\mathbf{a}_2+\boldsymbol{\delta}$, $\mathbf{R}_i+\mathbf{a}_1-\mathbf{a}_2+\boldsymbol{\delta}$, and 
$\mathbf{R}_i+\boldsymbol{\delta}$, respectively~\cite{Pujari2015}. Note that, since the variational wave function explicitly breaks translation 
symmetry, the order parameter (and not its square) can be directly assessed in the numerical calculation. For continuous transitions, we have:
\begin{eqnarray}
m^2 L^{1+\eta_m} &=& F_m \left [ \left(\frac{J_2-J_{cm}}{J_{cm}}\right) L^{1/\nu_m} \right ],\\
|\psi|^2 L^{1+\eta_p} &=& F_p \left [ \left(\frac{J_2-J_{cp}}{J_{cp}}\right) L^{1/\nu_p} \right ],
\end{eqnarray}
where $\nu_m$ ($\nu_p$) is the exponent for the magnetic (plaquette) correlation length, $\eta_m$ ($\eta_p$) is the exponent for this correlation 
function at criticality, and $J_{cm}$ and $J_{cp}$ are the values of $J_2$ at the transition points. Finally, $F_m$ and $F_p$ are suitable scaling 
functions. In the case of deconfined criticality, we must have $J_{cm}=J_{cp}$ and $\nu_m=\nu_p$, while the exponents are different, i.e., 
$\eta_m \ne \eta_p$. The results for the magnetization $m^2$ and for the plaquette order $|\psi|^2$ are reported in Fig.~\ref{fig:FSS}. Performing 
two separate fitting procedures based on a Bayesian statistical analysis~\cite{Harada}, we get $J_{cm}=0.234(1)$, $\nu_m=0.664(1)$, $\eta_m=0.837(1)$
for the magnetization, and $J_{cp}=0.224(1)$, $\nu_p=1.077(1)$, $\eta_p=0.799(1)$ for the plaquette order. These fitting procedures give a remarkably
good collapse of the two curves. Note that the evaluations of the critical points are in very good agreement, and also the values of $\eta_s$ and 
$\eta_p$ may be compatible with the prediction of the theory~\cite{Kaul2013}. However, the values of the exponents $\nu_m$ and $\nu_p$ are quite 
different, with an anomalously large value obtained for $|\psi|^2$. In fact, when attempting to fit both curves with the same $\nu$, a much worse 
result is obtained (not shown) and the data collapsing procedure fails in a large part of the magnetization curve.

When analyzing these scaling results, one must keep in mind that they are obtained within a variational approach, which may miss subtle details 
of the final phase diagram. Therefore, it can be very difficult to detect the existence of a deconfined quantum criticality. Nevertheless, 
it is striking that the two transitions look continuous with critical values $J_c$ that are extremely close to each other. The failure to obtain 
a good collapse with a single exponent $\nu$ could be due to the approximate nature of the variational wave function, which may not be particularly 
accurate in the VBS region (see Fig.~\ref{fig:accuracy}).

\section{Conclusions}\label{sec:conc}

In conclusion, we have employed variational wave functions and quantum Monte Carlo methods to study the frustrated $J_1\text{-}J_2$ Heisenberg 
model on the honeycomb lattice. We find that quantum fluctuations enlarge the region of stability of the collinear N\'eel phase with respect to 
the classical model, up to $J_2/J_1 \approx 0.23$. Further increasing $J_2/J_1$, a plaquette VBS order is stabilized, even though a gapless $Z_2$ 
spin liquid (dubbed $d\pm id$) represents a state with highly competitive variational energy, especially in the proximity of the phase transition.
We expect that this interesting new spin liquid can possibly be favored by farther-range couplings or by ring-exchange terms. 
At $J_2/J_1 \approx 0.36$, another VBS state with columnar order becomes energetically favored. Our results are in excellent agreement with recent 
DMRG calculations~\cite{Zhu2013,Gong2013,Ganesh2013b}.

Regarding the nature of the N\'eel to VBS transition, we hope that the promising results obtained by our approach will give a new impetus to 
examine the topic of a deconfined quantum critical point in the frustrated Heisenberg model on the honeycomb lattice.

\acknowledgements
We thank A.~Parola for providing us with the exact results on $24$ sites, and S.~Sorella for many useful discussions.
S.B. and F.B. thank R.~Thomale and Y.~Iqbal for useful discussions and for their hospitality at the University of W\"urzburg.
S.B. acknowledges the hospitality of SISSA and helpful conversations with C. Lhuillier.

\appendix

\section{PSG of the $d\pm id$ spin liquid}\label{app:PSG}

In this appendix, we shortly discuss the projective symmetry group~\cite{Wen2002,Bieri2016} and some physical properties of the competitive $d\pm id$
state discussed in this paper. For this purpose, we introduce a different formulation of the Hamiltonian of Eq.~\eqref{eq:hSL}, dropping the on-site 
terms which are not relevant for the present discussion:
\begin{equation}
{\cal H}_{\rm sl} = \sum_{i,j}  \psi^\dagger_i u_{ij} \psi_j =
\sum_{i,j} (c_{i,\uparrow}^\dagger,c_{i,\downarrow}^\dagga)  
\begin{pmatrix}
        t_{ij} & \Delta^*_{ij}  \\
        \Delta_{ij} & -t_{ij}^* \\
\end{pmatrix}
\begin{pmatrix}
        c_{j,\uparrow} \\
        c_{j,\downarrow}^\dagger \\
\end{pmatrix}.
\end{equation}
A spin liquid {\it Ansatz} $u_{ij}$ is invariant under the combined effect of a lattice symmetry transformation ($\mathcal{O}$) and the corresponding
gauge transformation ($g_\mathcal{O}$), namely
\begin{equation}
u_{ij}=g_\mathcal{O}(i) u_{\mathcal{O}^{-1}(i)\mathcal{O}^{-1}(j)} g_\mathcal{O}^\dagger(j).
\end{equation}

The $d\pm i d$ spin-liquid state is classified as No.~18 in Table~I of Ref.~\cite{Lu2011}. In the gauge employed in that paper, the quadratic 
mean-field Hamiltonian has a large cell (i.e., 6 sites). Here, we use a more natural gauge where the unit cell of the honeycomb lattice is not 
enlarged. In our gauge, both first-neighbor hopping phases and second-neighbor pairing phases undergo a $l=2$ phase winding as shown in 
Fig.~\ref{fig:ans18}. As a tradeoff for the simplicity of the {\it Ansatz}, the projective representation of symmetries is slightly more involved 
in this gauge.

\begin{figure}
\includegraphics[width=0.8\columnwidth]{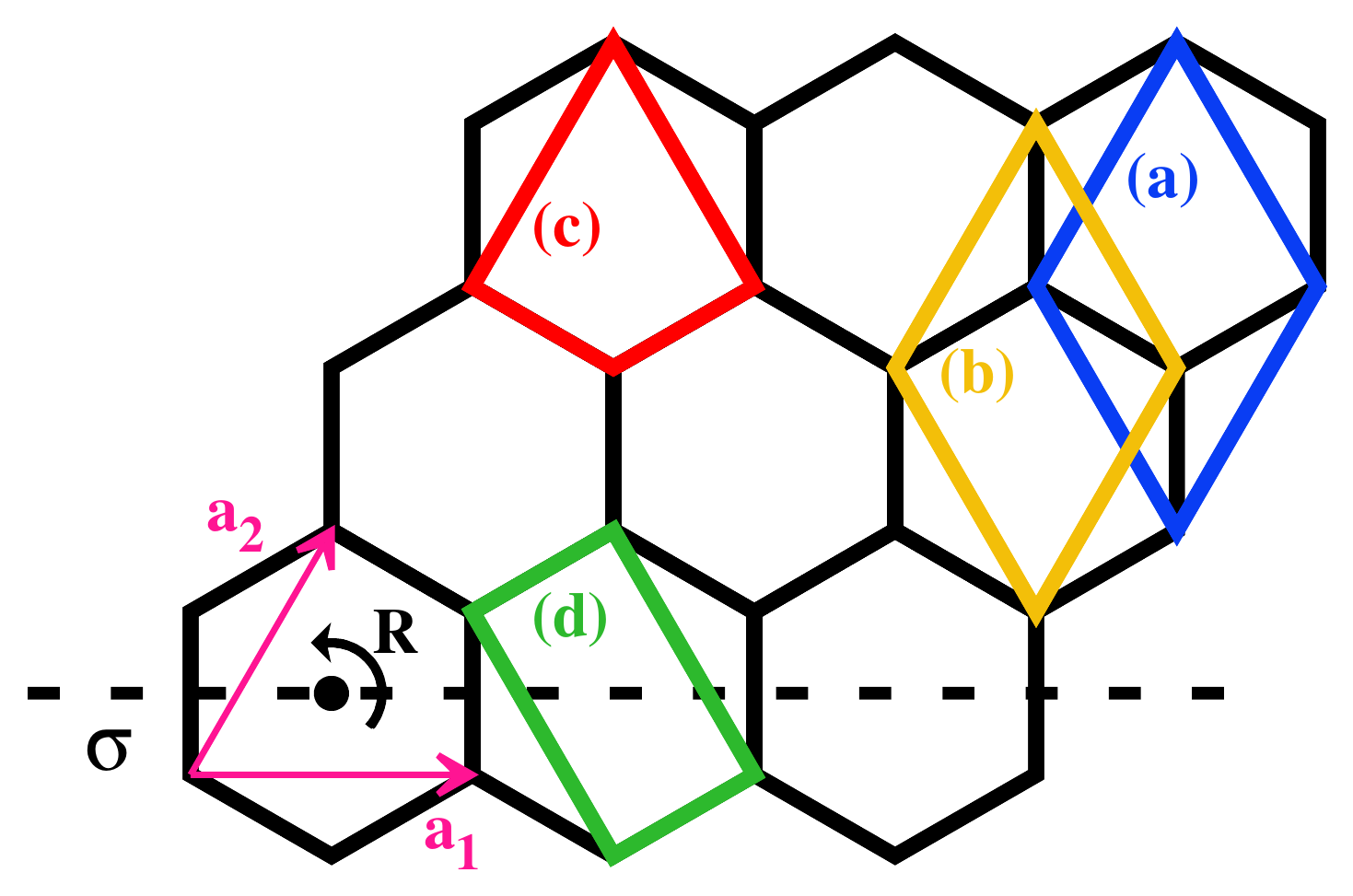}
\caption{\label{fig:fluxes}
Symmetry generators of the point group of the honeycomb lattice ($\sigma$ and $R$). As an example, we show the loops for which we computed the 
$SU(2)$ flux: (a) parallelogram-shaped (sublattice A), (b) parallelogram-shaped (sublattice B), (c) diamond and (d) rectangular plaquettes.}
\end{figure}

More explicitly, in our gauge we have trivial representations of lattice translations along $\mathbf{a}_1$ and $\mathbf{a}_2$, namely
$g_{1} = g_{2} = \mathbbm{1}_2$. The projective representation of the point group symmetries, i.e. the mirror reflection $\sigma$ and
the $6$-fold rotation $R$ (see Fig.~\ref{fig:fluxes}), is the following:
\begin{equation}\label{eq:gS}
g_\sigma(s) = \mathbbm{1}_2,
\end{equation}
\begin{equation}\label{eq:gR}
g_R(s) = (-)^s i\sigma_1 e^{(1-2s) \frac{\pi}{3} i \sigma_3},
\end{equation}
where $s=0,1$ is the sublattice index. Finally, for the time reversal $T$, we have
\begin{equation}\label{eq:gT}
g_T(s) = (-)^s i\sigma_3.
\end{equation}
For example, Eq.~(\ref{eq:gT}) implies that complex hopping terms between sites of different sublattices, and complex pairing terms between sites 
of the same sublattice are time-reversal invariant~\cite{Bieri2016}. The gauge transformation that relates our gauge for the $d \pm id$ state with 
the one used in Ref.~\cite{Lu2011} (No.~18 in Table~I) is given by:
\begin{equation}
g(j) =(-i\sigma_3)^s \exp\left[{i\frac{\pi}{12}\sigma_3}\right] \exp\left[{-i\frac{2\pi}{3}(n_j-m_j)\sigma_3}\right].
\end{equation}
In our gauge the {\it Ansatz} matrix reads
\begin{equation}
u_{ij}=
\begin{cases}
         t \sigma_3 \exp\left[i\phi_{ij}\sigma_3\right], \  (i,j) \mbox{ first-neighbor} \\
         \Delta \sigma_1 \exp\left[i\theta_{ij}\sigma_3\right], \  (i,j)  \mbox{ second-neighbor},
\end{cases}
\end{equation}
where the phases $\phi_{ij}$ and $\theta_{ij}$ are the ones of Fig.~\ref{fig:ans18}.

To conclude, let us discuss the gauge-invariant fluxes of the $d\pm id$ state on the honeycomb lattice. For any lattice loop $\mathcal{C}$ with 
base site $j$, we can define the SU(2) flux
\begin{equation}
P_j=\prod_{\mathcal{C}} u_{kl}= u_{j j_2} u_{j_2 j_3}\cdots u_{j_p j}.
\end{equation}
where $p$ is the number of sites in the loop. The trace of the $2\times2$ matrix $P_j$ is independent of the base site $j$~\cite{Bieri2016}
\begin{equation}
\Tr P_j=
\begin{cases}
         2\rho \cos(\theta), \  p \mbox{ even}\\
         2i\rho \sin(\theta), \  p \mbox{ odd},
\end{cases}
\end{equation}
and the angle $\theta$ is the gauge-invariant quantity that characterizes the SU(2) gauge flux.

For the $d \pm id$ {\it Ansatz}, pure first-neighbor loops have trivial fluxes ($\theta=0$), since the first-neighbor hopping is gauge equivalent 
to the uRVB state. The second-neighbor pairings, however, have nontrivial SU(2) flux with $\theta=\pm 2\pi/3$ through the parallelogram-shaped 
plaquettes of the triangular sublattices $A$ [Fig.~\ref{fig:fluxes}, loop (a)] and $B$ [Fig.~\ref{fig:fluxes}, loop (b)]. Odd-site loops do not 
contain nontrivial flux since the state is time-reversal invariant. As far as loops made from two first- and two second-neighbor links are concerned,
we can either have a diamond $J_1 J_1 J_2 J_2$ [Fig.~\ref{fig:fluxes}, loop (c)] or a rectangular $J_1 J_2 J_1 J_2$ [Fig.~\ref{fig:fluxes}, loop (d)]
plaquette. In the $d \pm id$ state, the trace of flux through the diamond plaquettes is trivial, while it gives $\theta=\pi$ through the rectangular 
plaquettes. These gauge-invariant fluxes are related to expectation values of certain multiple-spin operators~\cite{Bieri2016}.

\bibliographystyle{apsrev4-1}

\end{document}